# Kinetics of the Thermal Degradation of *Erica Arborea* by DSC: Hybrid Kinetic Method

D. Cancellieri[*], E. Leoni[*], J.L. Rossi.

SPE-CNRS UMR 6134 University of Corsica

Campus Grossetti B.P 52

20250 Corti (FRANCE).

* : corresponding authors

**Mail :** cancelie@univ-corse.fr, eleoni@univ-corse.fr

**Tel :** +33-495-450-076

**Fax :** +33-495-450-162

## Abstract

The scope of this work was the determination of kinetic parameters of the thermal oxidative degradation of a Mediterranean scrub using a hybrid method developed at the laboratory. DSC and TGA were used in this study under air sweeping to record oxidative reactions. Two dominating and overlapped exothermic peaks were recorded in DSC and individualized using a experimental and numerical separation. This first stage allowed obtaining the enthalpy variation of each exothermic phenomenon. In a second time, a Model Free Method was applied on each isolated curve to determine the apparent activation energies. A reactional kinetic scheme was proposed for the global exotherm composed of two independent and consecutive reactions. In fine mean values of enthalpy variation and apparent activation energy previously determined were injected in a Model Fitting Method to obtain the reaction order and the preexponential factor of each oxidative reaction. We plan to use these data in a sub-model to be integrated in a wildland fire spread model.

**Abbreviations:**

$E_a$: activation energy (kJ/mol)

$K_0$: preexponential factor (1/s)

$n$: reaction order

$T$: temperature (K)

$\beta$: heating rate (K/min)

$a_0, a_1, a_2, a_3$: numerical parameters of the interpolation function

$x, y, z$: interpolation coefficients

$I$: interpolation function

$r$: correlation coefficient

$\alpha$: conversion degree

$t$: time (min)

$R$: gas constant 8.314 kJ/mol

$f(\alpha)$: kinetic model reaction

$RSS$: residual sum of squares

$\Delta H$: enthalpy of the reaction (endo up) (kJ/g)

$\Delta m$: variation of mass loss (%)

$A$: virgin fuel

$KL$: reaction rate of char and volatiles formation

$B$: evolved gases

$K1$: reaction rate of reaction one

$B'$: oxidation products

$C$: chars

$K2$: reaction rate of reaction two

$D$: ashes.

**Subscripts:**

melt: melting point

Cur: Curie point

exo: exothermic

1: refers to exotherm 1

2: refers to exotherm 2

exp: experimental value

sim: simulated data.

offset: end of phenomena

deduct: deducted from experiment by substraction

iso: experimentally isolated

inter: interpolated

## 1. Introduction

The effect of fire on Mediterranean ecosystems has been a research priority in ecological studies for several years. Nevertheless, in spite of considerable efforts in fire research, our ability to predict the impact of a fire is still limited, and this is partly due to the great variability of fire behaviour in different plant communities [1,2]. Flaming combustion of ligno-cellulosic fuels occurs when the volatile gaseous products from the thermal degradation ignite in the surrounding air. The heat released from combustion causes the ignition of adjacent unburned fuel. Therefore, the analysis of the thermal degradation of ligno-cellulosic fuels is decisive for wildland fire modelling and fuel hazard studies [3-5]. Physical fire spread models are based on a detailed description of physical and chemical mechanisms involved in fires. Since the pioneering work of Grishin [6] these models incorporate chemical kinetics for the thermal degradation of fuels. However, kinetic models need to be improved.

Thermal degradation of ligno-cellulosic fuels can be considered according to Figure 1:

**GRAPHIC1.eps**

We present hereafter the results obtained on *Erica arborea*, one of the most inflammable species in Mediterrannean area. DSC curves showing two overlapped exothermic peaks (Exotherm 1 and Exotherm 2) were recorded at different heating rates under air. The fuel mass loss was recorded using TGA as an additional technique in order to get some information about the reactionnal mechanism. There are only a few DSC studies in the literature concerning the thermal decomposition of ligno-cellulosic materials which is preferably followed by TGA [7-9]. We adapted the DSC in order to measure the heat flow released by natural fuels undergoing thermal decomposition and in this paper we present a method to separate the thermal events from the global recorded exotherm. The two overlapped peaks observed on the DSC curves were experimentally and numerically isolated prior to the kinetic study.

The knowledge of the kinetic triplet ($E_a$, $K_0$ and $n$) and the kinetic scheme could help us in predicting the rate of thermal degradation when the collection of experimental is impossible in classical thermal analysis (high heating rates encountered in fire conditions). The thermal degradation kinetics of *Erica arborea* was studied using a combination of two kinds of methods: free model methods and model-fitting method. A free model method was applied in a first time to calculate the apparent activation energy and in a second time we used this result as initial data in a model fitting method to obtain preexponential factor, reaction order for our defined kinetic scheme. Our hybrid kinetic method is based on 4 stages (A – D) presented in this paper.

## 2. Experimental and methods of calculation

2.1. Experimental

Plant material was collected from a natural mediterranean ecosystem situated far away from urban areas in order to prevent any pollution on the samples. *Cistus monspeliensis (CM)*, *Erica arborea (EA)*, *Arbutus unedo (AU)* and *Pinus pinaster (PP)* are representative species of the Corsican vegetation concerned by wildland fires. In the present work we chose to focus on the results obtained from *EA* samples. Naturally, the methodology developed hereafter is applicable to every ligno-cellulosic fuel.

Only small particles (< 5 mm) are considered in fire spread [10]. Also, leaves and twigs were mixed, sampled and oven-dried for 24 hours at *60°C* [11]. Dry samples were grounded and sieved to pass through a 1 mm mesh, then kept to the desiccator. The moisture content coming from self-rehydration was about 4 percent for all the samples.

We recorded the Heat Flow *vs.* temperature (emitted or absorbed) thanks to a power compensated DSC ( Perkin Elmer®, Pyris® 1)  and the mass loss *vs.* temperature thanks to a TGA 6 (Perkin Elmer®).

The DSC calibration was performed out using the melting point reference temperature and enthalpy reference of pure indium and zinc ($T_{melt}$ *(In) = 429.8 K*, $\Delta H_{melt}$*(In) =  28.5 J/g*, $T_{melt}$ *(Zn) = 692.8*, $\Delta H_{melt}$*(Zn) = 107.5 J/g)*.  Thermal degradation was investigated in the range *400-900 K* under dry air or nitrogen with a gas flow of *20 mL/min*. Samples around *5.0 mg ± 0.1 mg* were placed in an open aluminium crucible and an empty crucible was used as a reference. The error caused by weighting gives an error of *1.9 % to 3 % on $\Delta H_{exp}$*.

We adapted the DSC for thermal degradation studies by adding an exhaust cover disposed on the measuring cell (degradation gases escape and pressure do not increase in the furnaces). Several experiments were performed with different high  heating rates (*$\beta$ = 10-40 K/min*) in order to be closer to the wildland fires conditions. A significant variation between the heating rates (*$\Delta\beta$=10 K/min*) was very important for kinetics purpose.

The TGA calibration was performed using the Curie point of magnetic standards: perkalloy® and alumel ($T_{Cur}$ *(alumel) = 427.4 K, $T_{Cur}$ (perkalloy®) = 669.2 K*). Samples around *10.000 mg ± 0.005 mg* were placed in an open platinum crucible and the degradation was monitored in the same range of temperature and heating rates as in DSC experiments.

## 2.2. Thermal separation

An experimental separation is very useful to indicate the way for a numerical treatment. Thanks to the switching of the surrounding atmosphere in the DSC furnaces we were able to define two independent and successive reactional schemes. The experimental conditions have been modified in order to hide the first exothermic phenomenon. Figure 2 present the schematic procedure we used to isolate the two phenomena with two experimental steps. The samples were thermally degraded under nitrogen atmosphere (step 1) at different heating rates from *400 K* to *900 K*. Then the residual charcoal formed during the step 1 was used as a sample to be analyzed by DSC under air sweeping (step 2) with the same temperature range and heating rates as in step 1. Step 1 allowed to pyrolyze the fuels generating a char residue and volatiles which escaped in the surrounding non-oxidizing atmosphere.

**GRAPHIC2.eps**

## 2.3. Numerical separation

The mathematical interpolation performed with *Mathematica®* [12] gave equations describing the DSC curves. We fitted the global curve obtained under air with Eq. 1 and Eq. 2. In a previous work [11] we used an empirical equation, with five adjustable parameters for each peak of each fuel, allowing the description of miscellaneous peaks and improved on temperature programmed desorption [13]. The following functions allowed a better fitting with only one parameters $a_0$ for Eq. 1 and three parameters $a_1, a_2, a_3$ for Eq. 2. These parameters were constant for all the species and heating rates. *Mathematica®* determined interpolation coefficient *x*, *y* and *z* in Eq. 1 and Eq. 2. $T_{exo1}$, $T_{exo2}$ are the temperatures of exotherms 1 and 2, $T_{shoulder}$ is the temperature of a shoulder observed in exotherm 2 for all the species, $T_{exo1}$, $T_{exo2}$ and $T_{shoulder}$ were deducted using the derivative of the experimental plot. The function which interpolates the set of experimental points is sought on the basis of the Eq. (1) equation model [14] for exotherm 1:

$$I_1(T) = \sum_{i=1}^{5} x_i \cdot \exp\left(-a_0 \cdot i \cdot \left(T_{exo1} - T\right)^2\right) \qquad (1)$$

and Eq. (2) equation model [14] for exotherm 2:

$$I_2(T) = \sum_{i=1}^{5} y_i \cdot \exp\left(-a_1 \cdot i \cdot \left(T_{shoulder} - T\right)^2\right) +$$

$$\sum_{i=1}^{15} z_i \cdot \left[1 + \exp\left(-\frac{T + a_2 \ln(i \cdot a_3) - T_{exo2}}{a_2}\right)\right]^{(1-a_3)} \cdot (i \cdot a_3)^{i \cdot a_3} \cdot \quad (2)$$

$$(i \cdot a_3 + 1)^{(i \cdot a_3 + 1)} \cdot \exp\left(-\left(\frac{T + a_2 \ln(i \cdot a_3) - T_{exo2}}{a_2}\right)\right)$$

with: $a_0 = a_1 = 0.001$; $a_2 = 50$ and $a_3 = 0.04$ for all the experiments. We chose these functions because they are more robust than traditional function and avoids long compilation times. Values of parameters $a_o$, $a_3$ and *n* are arbitrary, $a_1$ is correlated to the peak top temperature and $a_3$ is correlated to the peak width.

In order to fit to a list of an experimental data, we use a *Mathematica®* function: *"Fit[fun,data,var]"*. The data use have the form *{{$T_1$, $I_1$}, {$T_2$, $I_2$}, ... }*. This function finds a least-squares fit to a list of this data as a linear combination of the functions *fun* of variable *var (T)*. So, the solutions have the form: *I(T)=$C_o$+$C_1$ fun+$C_2$ fun$^2$+......+$C_n$ fun$^n$* where the interpolation coefficients: *Co......Cn* are given by *Fit*. In our case, *Fit* provides the solution: $C_i$=$x_i$ with *xo=0* for Eq. 1. and $C_i$={$y_i$, $z_i$} with *yo=zo=0* for Eq. 2.

In order to quantify the performance of the modelling procedure, for each experimental curve and the corresponding calculated one, Pearson's correlation ($r$) was measured and constrained to lie between -1 and 1. Variables are said to be negatively correlated, uncorrelated or positively correlated at temperature coordinates given by the experimental points.

## 2.4. Kinetic study

We have combined two kind of kinetic methods: model free kinetics and model fitting kinetics.

Both are based on Eq. 3:

$$\frac{d\alpha}{dt} = K_0 \cdot \exp\left(-E_a \big/ RT\right) \cdot f(\alpha) \qquad (3)$$

Model free kinetics is based on an isoconversional method where the activation energy is a function of the conversion degree of a chemical reaction. For this work we chose the method of Kissinger-Akahira-Sunose (KAS) applied without any assumption concerning the kinetic model ($f(\alpha)$). The KAS method [15] simply consists of extending the Kissinger's method [16] to the conversion range *0.1-0.9*, it is based on Eq. (4):

$$\ln\left(\frac{\beta_i}{T_{jk}^2}\right) = \ln\left(\frac{K_{0\alpha}R}{E_{a\alpha}}\right) - \frac{E_{a\alpha}}{RT_{jk}} - \ln g(\alpha_k) \qquad (4)$$

where $E_{a\alpha}$ and $K_{0\alpha}$ are respectively the apparent activation energy and the pre-exponential factor at a given conversion degree $\alpha_k$, and the temperatures $T_{jk}$ are those which the conversion $\alpha_k$ is reached at a heating rate $\beta_j$. During a series of measurements the heating rate are $\beta = \beta_1 \ldots \beta_j \ldots$ The apparent activation energy was obtained from the slope of the linear plot of $\ln\left(\beta_i\big/T_{jk}^2\right)$ *vs.* $1\big/T_{jk}$ performed thanks to a Microsoft® Excel® spreadsheet developed for this purpose. Four heating rates (*10, 20, 30, 40 K/min*) were used.

Model fitting kinetics is based on the fitting of Eq. 3 to the experimental values of *dα/dt*. We used Fork® (CISP Ltd.) software which is provided for model fitting in isothermal or non-isothermal conditions. The resolution of ordinary differential equations was automatically performed by Fork® according to a powerful solver (Runge Kutta order 4 or Livermore Solver of Ordinary Differential Equation). The reaction model *f(α)* was determined among six models specified in the literature [17]. Four heating rates (*10, 20, 30, 40 K/min*) were used at the same time; the software fit one kinetic triplet and one reaction model valid for all the heating rates. Once the determination of the best kinetic models and optimization of the parameters were achieved, the Residual Sum of Squares between experimental and calculated values (*RSS*) indicated the acceptable "goodness of fit" from a statistical point of view. The results presented in section 3 concern only the optimum parameters (best *RSS* value).

## 2.4. Hybrid Kinetic Method

Our hybrid kinetic method was built on four successive stages (A – B) from experimental data to simulated data. In stage A we individualized the exothermic phenomena. With stage B we obtained initiation data thanks to a Model free method applied on each phenomenon. The model free results were used as an initialization of the model fitting method. In stage C we proposed a kinetic scheme with two oxidative reactions and a Model Fitting Method gave the reaction model and the kinetic parameters of each phenomenon. Stage D was devoted to the simulation compared to experimental data in order to validate the method.

The application of our method to the thermal decomposition of *Erica arborea* gave the following results presented stage by stage in the next section.

## 3. Results and discussion

Figure 3 shows the experimental DSC/TGA thermograms for an experiment performed at *β = 30 K/min*. In this section, figures present only plot obtained for one heating rate to but two exotherms are clearly visualized and associated with two mass loss for all the heating rates. We chose to present in this paper only the results obtained on *Erica Arborea* fuel but the shape of thermograms from others fuels are very similar.

**GRAPHIC3.eps**

Table 1 presents the experimental results on the global exotherm in the range *400 – 900 K*, values of enthalpy variation were obtained by numeric integration on the whole time domain and peak top temperatures were determined thanks to the values of the derivative experimental curve.

Table 2 presents the results from TGA measurement for the considered heating rates. The first mass loss is clearly higher (around *70%*) than the second (around *27%*). The maximum temperatures of mass loss were determined thanks to the derivative experimental plot.

During the first exothermic process the plant is pyrolysed in the temperature range *400 K-600 K*, contributing to the formation of char. Gases emission are visualized in TGA by a mass loss around *70%*. An oxidation of these gases is possible when the surrounding atmosphere selected is air, this phenomenon is represented in DSC by the first exothermic peak. The second exothermic process can be considered like a burning process and it is known as glowing combustion. The char forms ashes in the temperature range of *600 K-900 K* , TGA plots show a mass loss around *27%* and the second exothermic peak is recorded in DSC. Other authors gave the same ascription for exotherm 1 and exotherm 2 [18-20].

We present hereafter the results obtained with the application of our original approach, the scope is the reduction a multi-step process in several independent steps.

3.1 Thermal and Numerical separation: **-Stage A-**

*Thermal separation*

As shown in Fig.4, in the range *400-900 K*, only the second exotherm is visualized in step 2 for all the species because only the remaining char was oxidized. Thus this experimental separation of exotherm 1 and exotherm 2 was very helpful to isolate the oxidation of char but the heat released by the oxidation of evolved gases could not been recorded.

**GRAPHIC4.eps**

We can deduct the variation of enthalpy for the first process by subtraction:

$$\Delta H_{1\,deduct} = \Delta H_{exp} - \Delta H_{2\,iso} \qquad (5)$$

Table 4 shows the results of numeric integration of isolated curves for the second oxidation and the deduction of the enthalpy variation for the first one from Eq. 5.

The values of enthalpy variation are constant for each reaction of each plant. We were able to give a mean value for the enthalpy variation of the gases oxidation (exotherm 1) and for the oxidation of char (exotherm 2) for each species. The obtained values are close whatever the heating rate is.

*Numerical separation*

Thanks to the interpolation functions Eq. 1 and Eq. 2 experimental DSC curves were reconstructed (exotherm 1 and exotherm 2) for all the heating rates considered.

Once the exotherms were plotted, enthalpy variation of each reaction was calculated by numerical integration of the signal and the results are shown in Table 4.

It is important to notice that the values obtained from the numerical treatment were found to be very close to those obtained by the thermal separation (*cf.* Tab. 3 and Tab 4.). The thermal *i.e.* experimental separation can be considered as a validation of the numerical separation.

We can say that the energy released by the reaction referred to exotherm 2 is more important than the energy released by the reaction referred to exotherm 1. Actually, we found a mean value of *4.74 ± 0.14 kJ/g* for the enthalpy variation of reaction referred to exotherm 1 with the associated mean value of mass loss of *70 ± 7 %* whereas we found a mean value of *7.45 ± 0.25 kJ/g* for the enthalpy variation of reaction referred to exotherm 2 with an associated mean value of mass loss of *27 ±1 %*.

Fig. 5 is an example of experimental data compared to interpolated curves. For this experiment we obtained a value of *r = 0.9946*. For all the species investigated and heating rates used the Pearson's correlation coefficient was about this value which indicates a very good fit. In the paper we present only results on experiments driven at *β = 30 K/min* in order to focus on the method we applied.

**GRAPHIC5.eps**

3.2 Initiation and Prediction: **-Stage B-**

In this stage, we used an isoconversionnal method in order to get initiation parameters (mean value of $E_a$) and also to have an idea of the mechanistic behaviour. Actually, the shapes of the dependence of $E_a$ on $\alpha$ have been identified from simulated data for competing [21], independent [22], consecutive [23], reversible [24] reactions and as well as for reactions complicated by diffusion [25].

Figures 6 and 7 show the results of KAS method applied on exotherm 1 and exotherm 2. In the range *0.1 - 0.9*, the global appearance of $E_a$ confirms the fact that these two mechanisms are different.

Figure 6 shows that the values of $E_a$ increase from *90* to *110 kJ/mol* in the range *0.1-0.4*, while they are relatively constant (about *110 kJ/mol*) in the range *0.4-0.9*. We did not take into account this low variation and only the global shape of the dependence of $E_a$ on $\alpha$ has been identified. Nevertheless, for the first process the $E_a$ values are nearly constant ($\cong$ *100 kJ/mol*) in the range *0.1-0.9*. Thus, this exothermic reaction could be considered as a single step process.

**GRAPHIC6.eps**

For the totality of second process $E_a$ values decrease from *140* to *90 kJ/mol* with an exponential shape. This relatively important variation shows that the second exothermic process is probably a multi-step process. The convex shape visualized on Fig. 7 is indexed in the literature [26] like a typical oxidation in solid state. Such processes occur frequently in the degradation of charcoal that decomposes as chars →ashes + gas.

**GRAPHIC7.eps**

3.3 Modelisation: **-Stage C-**

Let's consider the following kinetics scheme developed according to the results obtained in stage A and B.

**GRAPHIC8.eps**

We selected several reaction models [17] but the best results were obtained with a classical *n-th* order reaction: $d\alpha/dt = K_0 \cdot \exp\left(-E_a/RT\right) \cdot (1-\alpha)^n$ for both exothermic processes and considering two independent reactions.

The first process is modelled as: $A_{(s)} \rightarrow C_{(s)} + B'_{(g)}$. The measured heat flow correspond to the oxidation of evolved volatiles (exothermic) in gaseous state ($B_{(g)} \rightarrow B'_{(g)}$). Thus we studied indirectly the kinetics of $A_{(s)} \rightarrow C_{(s)} + B'_{(g)}$ by the kinetics of $B_{(g)} \rightarrow B'_{(g)}$ considering $A_{(s)} \rightarrow B_{(g)}$ as the rate limiting reaction of gas production. The second exothermic process concerns the oxidation of chars formed during the first process: $C_{(s)} \rightarrow D_{(s)} + E_{(g)}$.

The two processes can be traduced in the following differential equations considering a *n-th* order model:

$$\frac{d\alpha_1}{dT} = \frac{1}{\beta} \cdot K_{01} \cdot \exp\left(\frac{-E_{a1}}{RT}\right) \cdot (1-\alpha_1)^{n_1} \qquad (6)$$

$$\frac{d\alpha_2}{dT} = \frac{1}{\beta} \cdot K_{02} \cdot \exp\left(\frac{-E_{a2}}{RT}\right) \cdot (1-\alpha_2)^{n_2} \qquad (7)$$

Data (mean values) obtained in previous stage A and stage B are taken as initiation entries for the model fitting method in Fork® software (*cf.* Tab. 5) for a conversion degree varying from *0.1* to *0.9*.

The correlation coefficient and the *RSS* value helped us in identifying the "best" (statistically) set of parameters for a reaction model.

Table 6 presents the results obtained with the Model Fitting method, the parameters were calculated thanks to Fork® software and we present the best set of parameters for four heating rates (*10, 20, 30 and 40 K/min*). The first process of degradation has a coefficient of correlation $r_{exo1} = 0.9972$. In comparison the coefficient of correlation for the second process $r_{exo2} = 0,9941$, although satisfactory, indicates a lower value than $r_{exo1}$, it means that the *n-th* order model, would not correspond exactly to reality.

These results show that the mean values of $E_a$ and $\Delta H$ determined in the previous stages were very close to the values presented in Table 6. To calculate the parameters with Fork® the previous stages were absolutely necessary in order to avoid erroneous values and strong compensation effects in the set of parameters and $f(\alpha)$.

The results testify the coherent choice of the model and the reaction pathway selected. An unsuited model would have generated a divergence in the parameter determination. Fork® software allowed us to verify the validity of initial data ($E_a$ and $\Delta H$) calculated previously since the calculated parameters of Table 6 were in the ranges defined in Table 5.

3.4 Simulation and Prediction: **-Stage D-**

In this stage we used a formal resolution of Eq. 6 and Eq. 7 with Mathematica® software. The ordinary differential equation system was solved with the parameters calculated in stage C (Tab. 6). The resulting conversion degree was simulated on the whole temperature domain and allowed to plot the simulated heat flow. Figure 9 and 10 show the comparison between simulated plot and interpolated plot for exotherm 1 and exotherm 2. Figure 11 shows the comparison between the total simulated plot (sum of Eq. 6 and Eq. 7) and the global experimental plot.

**GRAPHIC9.eps**

**GRAPHIC10.eps**

**GRAPHIC11.eps**

The correlation coefficient shows the good fit of the simulated data to experimental ones. Of course we only present here the results obtained at one heating rate but simulations have been performed at different heating rates with similar correlation coefficients. We tested the validity of calculated parameters and *n-th* order models, simulations matched all our experimental data, we can simulate the thermal behaviour outside this range of temperature and heating rates, this will be presented in a future work.

## 4. Conclusion

As a general rule, reactions of thermal degradation show multi-step characteristics. They can involve several processes with different activation energies and mechanisms. On one hand, Model Free Kinetics considering only one kinetic parameter, namely $E_a$ is an over simplification of reality but they eliminate errors caused by an inappropriate kinetic model $f(\alpha)$. The validity of approaches, considering exclusively the activation energy values for the determination of the kinetics of solid state reactions, can be hardly accepted [17]. On the other hand, even if Model Fitting Methods give the kinetic triplet, they must be used carefully.

Without an initiation step, results tend to produce highly ambiguous kinetic descriptions [17]. We have developed an original method combining the previous ones to study the kinetics of thermal degradation reactions. We applied this hybrid kinetic method to the thermal degradation of ligno-cellulosic fuels encountered in wildland fires. DSC experiments on such fuels showed two superposed exothermic phenomena. In a first step, we individualized (numerically and experimentally) these phenomena in two oxidative sub-reactions. Enthalpy variations of each sub-reaction were calculated. After this simplification the activation energies for exotherm1 and exotherm2 were determined using a Model Free Method. Values of enthalpy variations and activation energies were used as initial data for the Model Fitting Method. In this way only two parameters of the kinetic triplet have to be determinate. We proposed a kinetic scheme for the thermal degradation of ligno-cellulosic fuels under air with nth-order model for the oxidative subreactions observed in DSC. Solid state kinetics computations should be carried out with experimental data obtained at several heating rates (not less than three) to ensure reliable results so, we performed this study with four heating rates (*10, 20, 30, 40 K/min*)  The application of these techniques is widely recognized for the characterization of the degradation of solids [17]. Simulations obtained with the calculated parameters and considering two *n-th* order reactions showed good fits to the experimental data. Such a method is promising to simulate heat flows at very high heating rates (outside the technical limits of classical TA).

In the field of wildland fires researches this approach could be useful since the step of thermal degradation of ligno-cellulosic fuels is still unknown. A sub-model of thermal degradation will be developed from this study and incorporated in a global physico-chemical fire spread model.

**References**

[1] F.A. Albini, Comb. Sci. Tech 42 (1985) 229.

[2] M. De Luis, M.J. Baeza, J. Raventos, Int. J. Wildland Fire 13 (2004) 79.

[3] A.P. Dimitrakopoulos, J. Anal. Appl. Pyrolysis 60 (2001) 123.

[4] R. Alèn, E. Kuoppala, P.J. Oesch, Anal. Appl. Pyrolysis 36 (1996) 137.

[5] J.H. Balbi, P.A. Santoni, J.L. Dupuy, Int. J. Wildland Fire 9 (2000) 275.

[6] A.M. Grishin, A.D. Gruzin, V.G. Zverev, Sov. Phys. Dokl. 28 (1983) 328.

[7] J. Kaloustian, A.M. Pauli, J. Pastor, J. Thermal Anal. 46 (1996) 1349.

[8] C.A. Koufopanos, G. Maschio, A. Lucchesi, Can. J. Chem. Eng. 67 (1989) 75.

[9] S. Liodakis, D. Barkirtzis, A.P. Dimitrakopoulos, Thermochim. Acta 390 (2002) 83.

[10] P. Caramelle and A. Clement  Res. Report. INRA, Avignon, 1978.

[11] E. Leoni, D. Cancellieri, N. Balbi, P. Tomi, A.F. Bernardini, J. Kaloustian, T. Marcelli, J. Fire Sci. 21 (2003) 117.

[12] T.B. Bahder, Mathematica for scientists and Engineers, Addison-Wesley, Reading, 1995.

[13] R. Spinicci, Thermochim. Acta 296 (1997) 87.

[14] M. Abramowitz, I.A. Stegun, Handbook of mathematical functions, Dover, New York, 1965.

[15] T. Akahira and T. Sunose, Res. Report. CHIBA Inst. Technol. 16 (1971) 22.

[16] H.E. Kissinger, Anal. Chem. 29 (1957) 1702.

[17] S. Vyazovkin, C.A. Wight, Int. Rev. Phys. Chem. 48 (1997) 125.

[18] J. Kaloustian, T.F. El-Moselhy, H. Portugal, Thermochim. Acta 401 (2003) 77.

[19] M.J. Safi, I.M. Mishra, B. Prasad, Thermochim. Acta 412 (2004) 155.

[20] C. Branca, C. Di Blasi, J. Anal. Appl. Pyrolysis 67 (2003) 207.

[21] S. Vyazovkin, A. Lesnikovich, Thermochim. Acta 165 (1990) 273.

[22] S. Vyazovkin, V. Goryachko, Thermochim. Acta 197 (1992) 41.

[23] S. Vyazovkin, Thermochim. Acta 236 (1994) 1.

[24] S. Vyazovkin, W. Linert, Inter. J. chem. Kinet. 27 (1995) 73.

[25] S. Vyazovkin, Thermochim. Acta 223 (1993) 201.

[26] S. Vyazovkin, C.A. Wight, Thermochim. Acta 340 (1999) 53.

**Appendix**

The following figures show the results of peak separation for experiments driven at 10, 20 and 40 K/min. One can also visualize the good fit of simulated data to experimental data for all the heating rates with the same set of parameters defined in Table 6.

**A1.eps**

**A2.eps**

**A3.eps**

**A4.eps**

**A5.eps**

**A6.eps**

**A7.eps**

**A8.eps**

**A9.eps**

**A10.eps**

**A11.eps**

**A12.eps**

**Acknowledgements**

The authors express their gratitude to the autonomous region of Corsica for sponsoring the present work. This research was also supported by the European Economic Community. We are pleased to acknowledge V. Leroy for having actively participated to the experiments.

**FIGURES LEGENDS**

Figure 1: Oxidative thermal decomposition of a ligno-cellulosic fuel.

Figure 2: Schematic representation of the experimental separation in two steps.

Figure 3: DSC/TGA curves for *EA* fuel at *30 K/min* under air sweeping.

Figure 4: DSC curve of residual char formed in step 1 at *30 K/min* under air sweeping.

**TABLES**

Table 1: Peaks temperature at different heating rates for EA fuel by DSC.

| $\beta$ (K/min) | $^a\Delta H_{exp}$ (kJ/g) | $^bT_{exo1}$ (K) | $^bT_{exo2}$ (K) |
|---|---|---|---|
| 10 | 12.11 | 624 | 769 |
| 20 | 12.61 | 649 | 789 |
| 30 | 12.13 | 666 | 805 |
| 40 | 12.16 | 678 | 820 |

[a] Total enthalpy of combustion calculated by numerical integration of the experimental signal on the whole temperature range.

[b] $T_{exo1}$ and $T_{exo2}$ are the temperature of maximum value of heat flow obtained by DSC measurements.

Table 2: Offset temperature and mass lost at different heating rates for EA fuel by TGA.

| $\beta$ (K/min) | $\Delta m_1$ (%) | $T_{offset1}^{a}$ (K) | $\Delta m_2$ (%) | $T_{offset2}^{a}$ (K) |
|---|---|---|---|---|
| 10 | 68.3 | 644 | 27.4 | 796 |
| 20 | 69.4 | 660 | 28 | 808 |
| 30 | 77 | 676 | 26.6 | 819 |
| 40 | 66.2 | 693 | 28.6 | 832 |

[a] $T_{offset1}$ and $T_{offset2}$ are the temperature of maximum value of mass loss obtained by TGA measurements.

Table 3: Enthalpies variations of reaction 1 deducted and reaction 2 isolated.

| $\beta$ (K/min) | $\Delta H_{1deduct}$ (kJ/g) | $\Delta H_{2\,iso}$ (kJ/g) |
|---|---|---|
| 10 | 4.59 | 7.52 |
| 20 | 5.14 | 7.47 |
| 30 | 4.70 | 7.43 |
| 40 | 4.71 | 7.45 |

Table 4: Enthalpies variations of reaction 1 interpolated and reaction 2 interpolated.

| $\beta$ (K/min) | $\Delta H_{1\,inter}$ (kJ/g) | $\Delta H_{2\,inter}$ (kJ/g) |
|---|---|---|
| 10 | 4.62 | 7.35 |
| 20 | 4.88 | 7.70 |
| 30 | 4.73 | 7.38 |
| 40 | 4.74 | 7.39 |

Table 5: Initiation data obtained in stage A and stage B.

| Initial data | | |
|---|---|---|
| Exotherm 1 | $4.62 \leq |\Delta H_1|$ (kJ/g) $\leq 5.38$ | $80 \leq E_{a1}$ (kJ/mol) $\leq 120$ |
| Exotherm 2 | $7.35 \leq |\Delta H_2|$ (kJ/g) $\leq 8.20$ | $80 \leq E_{a2}$ (kJ/mol) $\leq 150$ |

Table 6: Summary of kinetics results (Model Fitting) obtained for exotherm 1 and exotherm 2.

| Parameters | Results$_{1 \text{ inter}}$ | Results$_{2 \text{ inter}}$ |
| --- | --- | --- |
| $\ln K_o$ | 13.2 | 12.9 |
| $E_a$ | 95.3 kJ/mol | 116.3 kJ/mol |
| $n$ | 1.68 | 0.49 |
| $|\Delta H|$ | 4.98 kJ/g | 7.56 kJ/g |
| $r_{exo}$ | 0.9972 | 0.9941 |